\documentclass[aps,pra,preprint,groupedaddress]{revtex4}
\usepackage{color}
\usepackage{epsf}
\usepackage{epsfig}
\usepackage{graphicx}
\usepackage{latexsym}
\usepackage{bm}
\usepackage{amssymb}

\begin{document}

\title{2D electron momentum distributions for transfer ionization in fast proton Helium collisions}

\author{M.~S.~Sch\"{o}ffler$^{1}$}
\email{schoeffler@atom.uni-frankfurt.de}
\author{O.~Chuluunbaatar$^{2,3}$}
\author{S. Houamer$^4$}
\author{A. Galstyan$^5$}
\author{J.~N.~Titze$^1$}
\author{L.~Ph.~H.~Schmidt$^1$}
\author{T.~Jahnke$^1$}
\author{H.~Schmidt-B\"{o}cking$^1$}
\author{R.~D\"{o}rner$^1$}
\author{Yu.~V.~Popov$^6$}
\author{A.A. Gusev$^2$}
\author{C. Dal Cappello$^7$}

\affiliation{$^1$ Institut f\"ur Kernphysik, University Frankfurt,
Max-von-Laue-Str. 1, 60438 Frankfurt, Germany}

\affiliation{$^2$Joint Institute for Nuclear Research, Dubna,
Moscow region 141980, Russia}

\affiliation{$^3$ School of Mathematics and Computer Science,
National University of Mongolia, UlaanBaatar, Mongolia}

\affiliation{$^4$D\'{e}partement de physique, Facult\'{e} des
Sciences, Universit\'{e} Ferhat Abbas, S\'{e}tif, 19000, Algeria}

\affiliation{$^5$Faculty of Physics, Lomonosov Moscow State
University, Moscow 119991, Russia}

\affiliation{$^6$Skobeltsyn Institute of Nuclear Physics,
Lomonosov Moscow State University, Moscow 119991, Russia}

\affiliation{$^7$Institut de Chimie, Physique et des Mat\'eriaux,
Universit\'e de Lorraine 1 Bd Arago 57078 METZ Cedex 3, France}

\date{\today}

\begin{abstract}
The momentum distribution of the electron in the reaction p+He
$\rightarrow$ H + He$^{2+}$ + $e$ is measured for projectile
energies $E_p$=300 and 630 keV/u at very small scattering angles
of hydrogen. We mainly present two dimensional distributions
parallel $(k_{||})$ and perpendicular $(k_{\perp})$ to the
projectile beam. Theoretical calculations were carried out within
the Plane Wave First Born Approximation (PWFBA), which includes
both electron emission mechanisms, shake-off and sequential
capture and ionization. It is shown that electron correlations in
the target wave function play the most important role in the
explanation of experimentally observed backward emission. Second
order effects have to be involved to correctly describe the
forward emission of the electron.
\end{abstract}


\maketitle

\section{Introduction}

In the past decade a new wave of theoretical and experimental
interest in electron capture processes, involving two active
electrons, as double capture (DC), transfer ionization (TI) and
transfer excitation (TE), has shed light on the versatile effects
of electron correlation. New experimental techniques allow to
measure more than only total or single differential cross section
(SDCS). Fully differential cross sections (FDCS), which depend on
the momentum distribution of the escaped electron in TI give a
rather detailed view in the dynamical processes taking place. In
particular, in this paper we consider the reaction p+He
$\rightarrow$ H+He$^{2+}$+$e^-$.

Since the early publications \cite{OBK,Thomas} it became clear
that two principal mechanisms contribute to the transfer
ionization. This takes place via a capture of one electron with a
correlated (shake-off, SO) or sequential process (binary
encounter, BE), removal of the second electron. We use this
terminology in accord with single photon ionization of an atom
\cite{Briggs2000jpb, Knapp2002prl} in spite of quite different
transfer energies in both cases. Let us concentrate further on the
single transfer ionization, because this process is a subject of
this paper. Direct capture presumes the "usurpation" of one target
electron by the fast projectile proton, like it was described in
\cite{OBK}, and releasing of another electron due to the sudden
rearrangement of the field in the residual ion (typical SO). If
the fast proton is described by the plane wave in the lab frame,
and its scattering angles are very small (fractions of mrad), then
the OBK-mechanism \cite{OBK} presents the principal transition
matrix element alike to that for Electron Momentum Spectroscopy
\cite{EMS} (see also \cite{JETP}). In turn, it was shown that
latter one is very sensitive to angular and radial
electron-electron correlations in the target \cite{Takahashi}.

The captured electron always moves forward parallel to the
velocity vector of the proton projectile, i.e. its momentum
component is positive. If the electron-electron correlation in the
target is weak (say, only due to a mean field), the emitted
electron will be shaken off isotropically. In the opposite case of
strong angular correlations it moves predominantly in backward
direction ($k_{||}<0$) and we expect to see a backward peak in the
electron momentum distribution. A different process (analogue to
radiative electron capture) also resulting in a backward emitted
electron was suggested by Voitkiv and coworkers
\cite{Voitkiv2008prl,Schulz2012prl}. These calculations lack of
high differentiality, as the are neither in the scattering angle
dependent nor in the scattering plane. Therefore we will show our
data only in the longitudinal vs. transversal representation.

The sequential mechanism of TI presumes at least two successive
interactions of the fast projectile with both target electrons.
For its realization no electron-electron correlations are needed.
This mechanism in general is of the second order (and higher) in
the projectile-target interaction. However, features of the
capture processes allow to define transfer ionization already with
a first order amplitude \cite{Salim10}. After interaction of the
bound electron with the fast projectile proton it becomes also
fast. It can interact again with another electron or the target
nucleus on its way out (pure second order $Ne$- and $ee$-Thomas,
for example \cite{Thomas,Briggs}), but its movement keeps in
general the forward character $k_{||}>0$. So, the forward peak can
be connected with the BE mechanism; capture and ionization are
generally independent.

Of course, the above considerations are semiclassical, we shall
see an interplay of quantum mechanisms and coherent sum of
corresponding matrix elements, but we expect the general
forward-backward features to be present also in a full quantum
treatment.

We think, it is a time to defend the PWFBA, because first Born
theories are often believed to be inadequate for electron capture.
We would like to stress that it is not so. First, any FBA theory
works well until the higher Born terms become bigger in the region
of final state phase space considered. So for example at very
small scattering angles of a fast projectile ion (proton), the OBK
term is a leading one but if falls down rapidly with increasing
scattering angle, and the higher order terms begin to contribute.
But they do not contribute much at very small angles (see
calculations in \cite{Kim}). Second, the OBK matrix element, as it
was considered 80 years ago, now can include much better
correlated trial wave functions. This plays a crucial role for
transfer excitation and transfer ionization reactions (less for
charge transfer). At very small projectile scattering angles, the
corresponding SDCS curves for highly and loosely correlated target
ground functions start to differ substantially. Third, we have now
two main approaches for capture reaction: time-dependent
semiclassical and pure quantum mechanical. Within the
time-dependent approach the heavy fast projectile particle is
considered a source of a classical outer field. The interaction of
both, projectile and target ion, can easily be taken into account
as a phase-factor to the final (initial) wave function
\cite{Belcic}. In the pure quantum approach (see, for example,
\cite{Kim}) the interaction of heavy particles gives a
contribution to the FBA, and this term distorts noticeably the OBK
term both at very small (here the distortion is "positive", it
diminishes the peak value) and at larger scattering angles (here
this distortion is "negative", it increases the plateau). It was
shown that SBA terms can compensate this negative effect and
considerably improve the agreement between theory and experiment
\cite{Kim}. Physically it is clear that if the transferred
momentum and energy are relatively small (we are in a laboratory
frame, and this takes place at very small scattering angles), then
the velocity of atomic nucleus is practically zero, and it is
actually immovable during the scattering process. The nucleus
simply changes the initially directed path of the projectile due
to elastic scattering ("secondary" OBK). This is the main physical
role of this term at scattering angles close to zero. But its FBA
realization distorts this picture at larger angles, and the SBA
provides with necessary corrections.

In this paper we present experimental results and calculate fully
differential cross sections (FDCS) within the plane wave first
Born approximation (PWFBA) on proton-helium interaction at impact
energies of 300 and 630 keV. Both discussed above mechanisms, SO
and BE, contribute in this case.

Atomic units $\hslash = e = m_e = 1$ are used throughout unless
otherwise specified.

\section{Experiment}

To achieve the goals of this experiment all emitted particles have
to be measured in coincidence. Therefore we applied momentum
spectroscopy techniques, as reactions microscopes or COLTRIMS
(COLd Target Recoil Ion Momentum Spectroscopy)
\cite{Ullrich1997jpb,Doerner2000pr,Ullrich2003rpp}. The
experiments were performed at the Institut f\"ur Kernphysik at the
University of Frankfurt using the Van de Graaff accelerator. Using
3 sets of movable slits, the proton beam was collimated to a
divergence less than 0.15 mrad, an size of about 0.5 $\times$ 0.5
mm$^2$ at the overlap region with the gas jet. 15 cm upstream of
the target, a set of parallel electrostatic deflector plates
cleaned the primary beam from charge state impurities, deflecting
the primary beam slightly upwards. The H$^+$ beam was crossed
perpendicular with the helium gas jet. 15 cm downstream of the
target a second set of horizontal electrostatic deflector plates
separate the final charge state, thus only the neutral projectiles
H hit a position and time sensitive multichannel plate (MCP)
detector, placed 3 m downsteam the interaction point, yielding the
projectile deflection angle and the time zero of the collision.
The main part of the beam ($\approx$ 1 nA), which is still charged
was dumped in a Faraday cup.

The gas jet providing the target beam was generated by helium
gasexpanding through a 30 $\mu$m nozzle with a backing pressure of
20 bar and collimated in a two stage jet. A density of
$5\times10^{11}$ atoms/cm$^2$ and a diameter of 1.5 mm were
achieved. The active cooling by the supersonic expansion in
expansion direction combined with passive one in the perpendicular
direction by the geometry resulted in a 3 dimensional cold target
and a momentum uncertainty below 0.1 a.u.

At the intersection volume where proton and helium beam were
intersected, electrons and ions were created. A weak electrostatic
field of 4.8 V/cm was applied to project electrons and recoiling
ions onto two position and time sensitive detectors. To optimize
the resolution, a three dimensional time and space focusing
geometry \cite{Schoeffler2011njp,Mergel1995prl} was used for the
recoil ion arm of the spectrometer. The ion were detected by a
80~mm diameter micro channel plate (MCP) detector with delay-line
anode \cite{Jagutzki2002nima,Jagutzki2002nima2}. The time focusing
was realized using a field free drift tube \cite{WWil55}, while an
adjustable electrostatic lens was used to achieve space focusing.
This lens was optimized by minimizing the spatial width of the
lines on the detector from He$^+$ ions created by pure capture,
which have been recorded parallel to the transfer ionization
events (for an example see Fig. 1 in \cite{Schoeffler2009pra1} or
Fig. 1 in \cite{Kim}). A momentum resolution of 0.1 a.~u. was
achieved in all three directions. The electrons were guided by a
magnetic field (see \cite{Moshammer96nim}) of 15 and 25~Gauss and
accelerated over a length of 20~cm by the same electric field in a
time focusing geometry (40~cm additional field free drift tube)
onto a MCP detector of 120~mm active diameter. The overall
spectrometer geometry, especially the ion's part was simulated
using SIMION to gain the maximum resolution and efficiency.

We reached an overall acceptance of 4$\pi$ solid angle for recoil
ions up to a momentum of 10 a.u. and electrons up to 6 a.u. A
three-particle coincidence (H$^0$+He$^{2+}$+$e$) was applied to
record the data event by event. From the positions of impact on
the detectors and the time-of-flight we can derive the initial
momentum vectors of the recoil ion and the electron. The
projectile transverse momentum vectors were directly measured.
Checking energy and momentum conservation the background was
strongly suppressed during the off-line data analysis. Also the
overall resolution was good enough to measure the final electronic
state of the H and separate events where the hydrogen was found in
the ground state from where the electron was captured into an
excited state. Only these events, where the hydrogen is in the
ground state are presented in the following.

\section{Theory}

As stated above, we consider the He atom as a target for the TI
reaction. We follow definitions and notations given in
\cite{Salim10} and not repeat all conditions here. In the momentum
representation in the lab frame and at very small scattering angle
$\theta_p$ the symmetrized matrix element is given by
$$
\mathcal{T}_{FBA} = -4\pi\sqrt 2 \int \frac{d\vec x}{(2\pi)^3} \frac{
\widetilde{\phi}_{H}(x)} {\vert \vec v_p - \vec q - \vec
x\vert^2}[F(\vec q;0;\vec k) + F(\vec v_p - \vec x; -\vec v_p +
\vec q + \vec x; \vec k)
$$
$$
- 2F(\vec v_p - \vec x;0;\vec k)]=A1+A2+A3, \eqno (1)
$$
where
$$
{F(\vec y; \vec \eta; \vec k)} = \int e^{-i\vec y \vec r_1 - i\vec
\eta \vec r_2} \varphi_c^{-*}(\vec k, \vec r_2)\Phi_0 (\vec r_1,
\vec r_2) d\vec r_1 d\vec r_2, \eqno (2)
$$
$\vec v_p$ is the fast proton velocity, the transferred momentum
$\vec q=\vec p_H-\vec p_p$, $\vec k$ the electron momentum,
$\Phi_0({\vec r_1},{\vec r_2})$ the helium ground wave function,
and the Coulomb wave function of the final target ion
$$
\varphi_c^{-*}(\vec k, \vec r) =e^{-\pi\xi/2}\Gamma(1+i\xi)e^{-i{\vec k}{%
\vec r}}{_1}F_1(-i\xi,1;ikr+i{\vec k}{\vec r}); \quad \xi=-2/k.
$$

The FDCS is calculated by the formula
$$
\frac{d^2\sigma}{dk_{\perp}
dk_{||}}=\frac{m^2k_{\perp}}{(2\pi)^4}\int\limits_0^{\theta_{max}}\theta_pd\theta_p
\int\limits_0^{2\pi}d\phi_k |A1+A2+A3|^2, \eqno (3)
$$
with $m=1836.15$ being the proton mass. We display all vectors'
components for clarity: $\vec v_p=\{0,0,v_p\},\ \vec
q=\{mv_p\theta_p,0,q_{||}\},\ \vec
k=\{k_\perp\cos\phi_k,k_\perp\sin\phi_k,k_{||}\}$. We also remind
that $q_{||}={v_p}/{2}+{Q}/{v_p}$ with $Q=E_0^{He}-E^H-k^2/2$.

In (1) the term $A_1$ is the OBK amplitude, where any trial helium
wave function can be used. The amplitude $A_3$ can also be
attributed to SO. It describes the contribution of heavy particles
interaction and was discussed in the Introduction. The amplitude
$A_2$ is a typical PWFBA realization of the BE mechanism.

\section{Results and Discussion}

For calculations we use three trial helium wave functions. One is
the loosely correlated $1s^2$ Roothaan-Haartree-Fock (RHF)
function of Clementi and Roetti \cite{RHF} ($E_0^{He}$=-2.8617).
The two others are a highly correlated function of the type
$$
 \Psi(r_1,r_2,r_{12})=\sum_{j=1}^{N}D_j\left[\exp(-\alpha_j
 r_1-\beta_j r_2)+\exp(-\alpha_j r_2-\beta_j
 r_1)\right]\exp(-\gamma_j r_{12}), \eqno (4)
$$
which was described in \cite{Chuka06} ($E_0^{He}$=-2.9037), and
the configuration interaction (CI) wave function of Mitroy
\cite{Mitroy} ($E_0^{He}$=-2.9031).

\begin{figure}[htb]
\centering
\includegraphics[width=10cm]{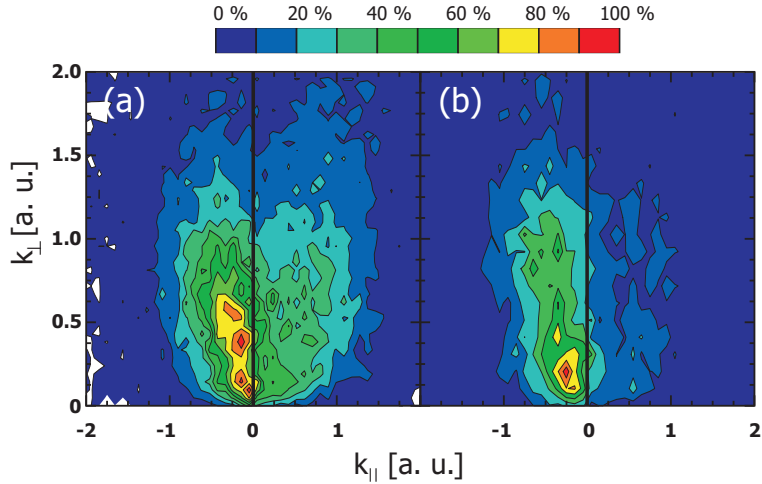}
\caption{(Color online) Experimental momentum distribution of the electron
for a) $E_p=$300 keV and b) $E_p=$630 keV. The projectile is moving in the
positive $k_{||}$ direction, i. e. from the left to the right.
The data are integrated over all other observables, i. e. the integral
over the shown distribution corresponds to the total transfer ionization
cross section for the H(n=1) state.}
\end{figure}

The experimental data at $E_p=300$ and $E_p=630$ keV, shown in
Figure 1, display a noticeable peak at backward (negative
$k_{||}$) direction and a less resolved peak at forward direction
(positive $k_{||}$). The forward peak structure has more intensity
at the lower projectile energy of 300 keV, as the
projectile-target interaction time is longer and therefore an
additional interaction, the electron knock-off, more likely to
occur.

As expected calculations with the loosely correlated wave function
[Figs. 2(c) and 2(d)] give practically no backward peak to the
electron's distribution. Both highly correlated helium wave
functions give very similar distributions [Figs. 2(a) and 2(b)],
which include both forward and backward peaks. However, visually
they are hard to compare with the experiment.

To avoid effects of color scales, we present additionally two
slices of these distributions at $E_p=300$ keV and fixed
$k_\perp$: $k_\perp$=0.2 in Fig. 3(a) and $k_\perp$=0.4 in Fig.
3(b). The experimental points are normalized to the theory's peak
maximum along the whole distributions. First, we clearly see that
both used correlated wave functions give practically the same
curves. Second, theory and experiment well coincide at negative
$k_{||}$, what clearly demonstrate that the PWFBA shake-off
amplitude is quite sufficient to describe the backward peak. This
requires of course, the use of highly correlated target wave
functions. Third, we see that the theory noticeably exceeds the
experimental points in the forward domain $k_{||}>0$. It is a
clear indication that the SBA calculations are needed here.
Unfortunately, we cannot provide these calculations at the moment.

We finally show a comparison of the total transfer ionization
cross section in PWFBA theory, using Mitroy helium wave function,
and experiment. In Figure 4 the agreement is quite satisfactory
over a wide range of the proton energies.

\section{Conclusions}

In conclusion, we presented highly differential theory (PWFBA) and
experimental data from a kinematical complete experiment on
transfer ionization in proton-Helium-collision at 300 and 630 keV.
The observed splitting into forward and backward emission
originates from two different contributions, the $A_2$-term
(binary encounter) and the $A_1+A_3$-term (shake-off). Comparison
of loosely and highly correlated wave functions for the initial
state confirms the high sensitivity of the experiment to the
subtle features of the initial state wave function. Better
agreement for the forward emitted electrons can be expected for
calculations in the second order. At the same time, backward
emitted electrons can be described within the first Born
approximation at high projectile energies.

\section{Acknowledgements}

We acknowledge financial support from the Deutsche
Forschungsgemeinschaft (DFG), Grant No. SCHO 1210/2-1. As well
this work partially supported by Russian Foundation of Basic
Research (RFBR), Grant No. 11-01-00523-a. All calculations were
performed using Moscow State University Research Computing Centre
(supercomputers Lomonosov and Chebyshev) and Joint Institute for
Nuclear Research Central Information and Computer Complex. The
authors are grateful to K. Kouzakov for inspiring discussions and help.

\begin{figure}[htb]
\centering
\includegraphics[width=12cm]{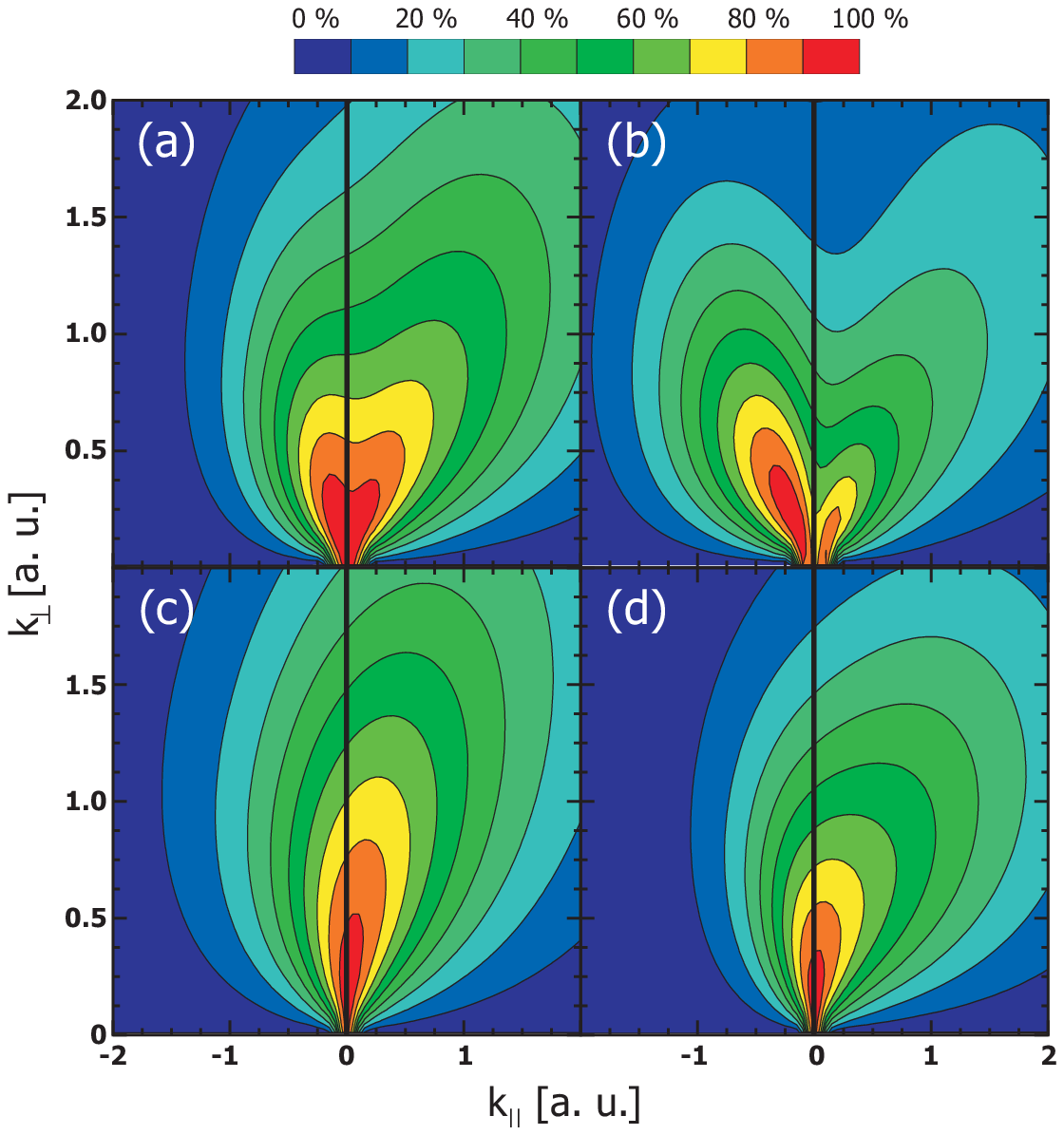}
\caption{Calculated electron momentum distribution longitudinal vs. transversal
within the PWFBA for a) $E_p=$300 keV with  strong correlation, b) $E_p=$630 keV
with strong correlation, c) $E_p=$300 keV with weak correlation, d) $E_p=$630 keV
with  weak correlation}
\end{figure}

\begin{figure}[htb]
\centering
\includegraphics[width=12cm]{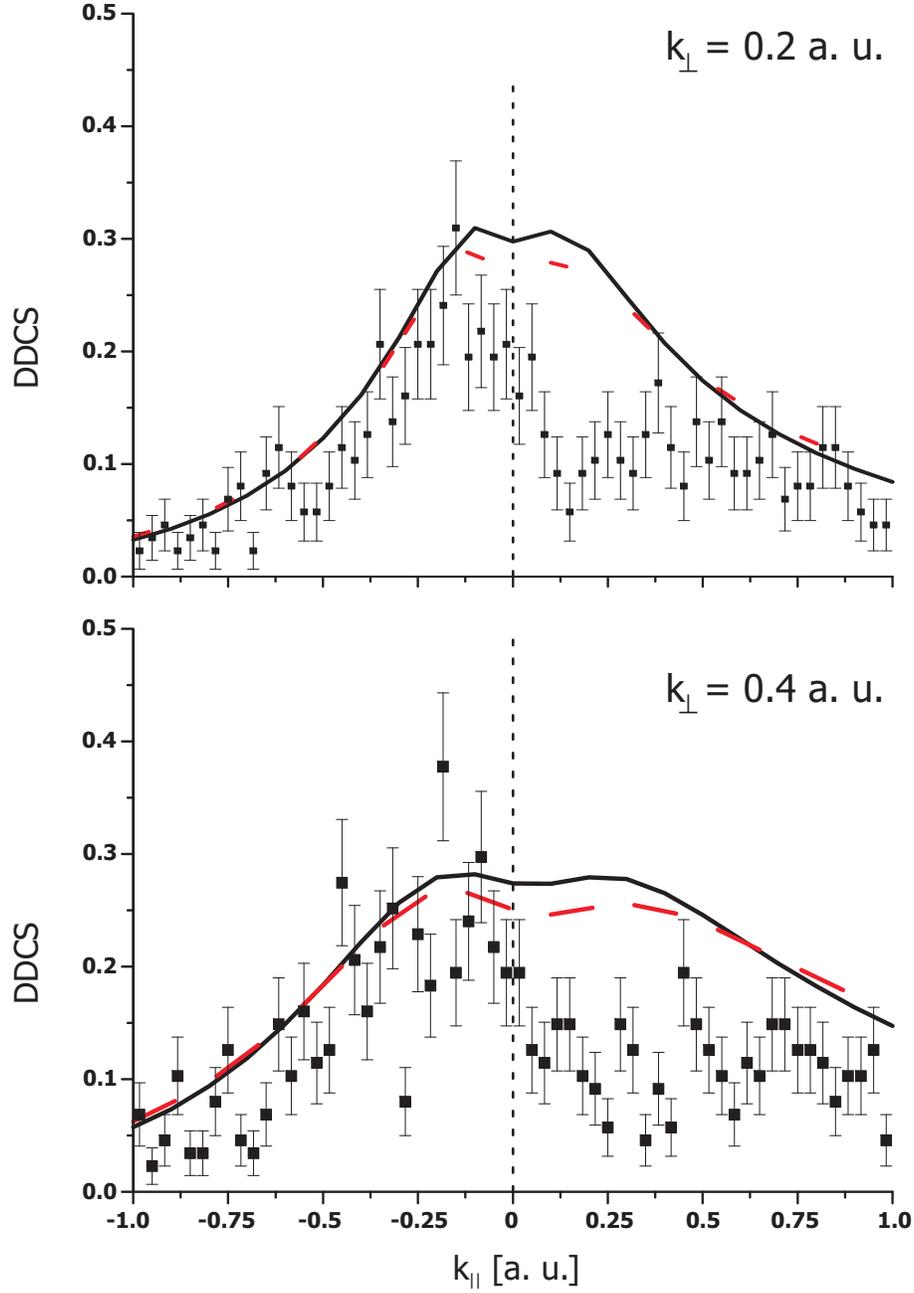}
\caption{DDCS versus $k_{||}$ at $E_p=300$ keV for fixed $k_\perp=0.2$ (top),
$k_\perp=0.4$ (bottom). Solid line, highly correlated wave function \cite{Chuka06};
dashed line, that of Mitroy. Dots are experimental points.}
\end{figure}

\begin{figure}[htb]
\centering
\includegraphics[width=\linewidth]{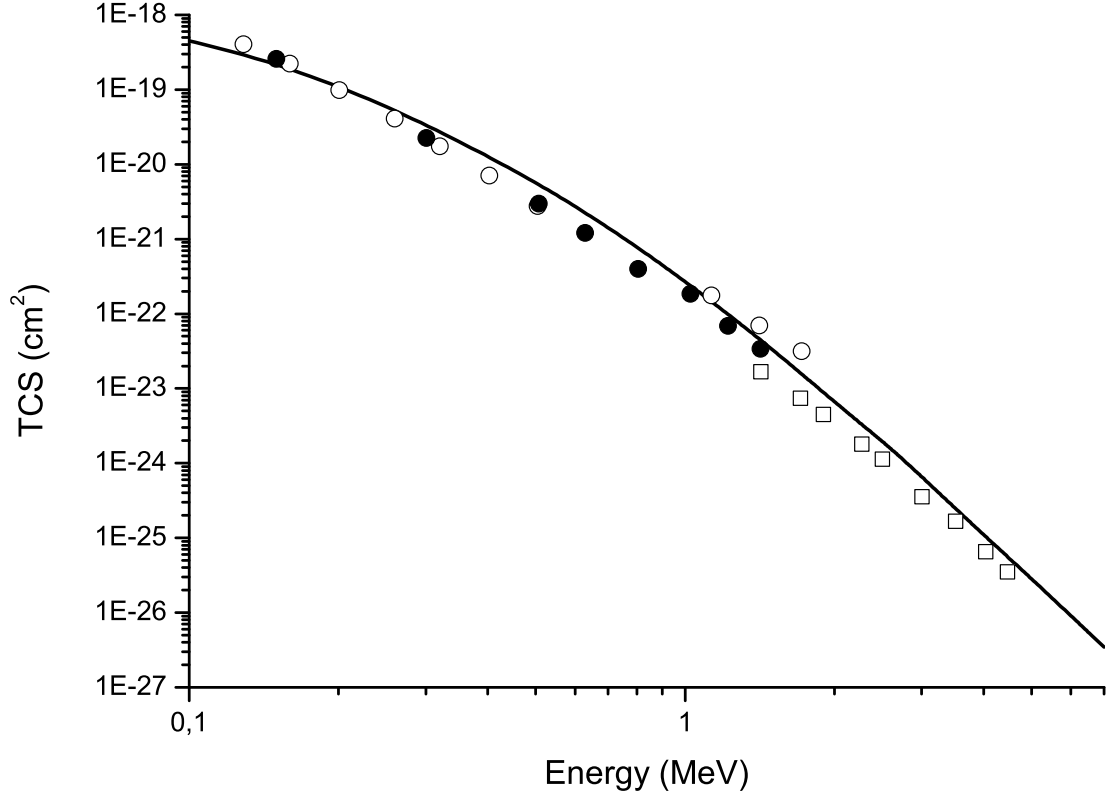}
\caption{Total cross section for transfer ionization for different proton energies
$E_p$(solid line), using Mitroy wave function. Experiment: open circles, Shah and Gilbody \cite{Shah};
full circles, Mergel et al. \cite{Mergel97}; open squares, Schmidt et al \cite{Schmidt05}}
\end{figure}

\end{document}